\documentclass{ENZCon}

\usepackage{graphicx} 

\begin{document}

\title{Current New Zealand Activities in Radio Astronomy: Building Capacity in Engineering \& Science for the Square Kilometre Array}

\author{M.Johnston-Hollitt\inst{1}, V. Kitaev\inst{2}, 
 C.P. Hollitt\inst{3}, N. Jones\inst{4}, T.C. Molteno\inst{5}\\
\institute{School of Chemical \& Physical Sciences, 
Victoria University of Wellington\\
Box 600, Wellington, NEW ZEALAND. \\
Email: Melanie.Johnston-Hollitt@vuw.ac.nz}\\
\institute{School of Engineering, Auckland University of Technology\\
Private Bag 92006, Auckland, NEW ZEALAND.} \\
\institute{School of Engineering \& Computer Science, 
Victoria University of Wellington\\
Box 600, Wellington, NEW ZEALAND.} \\
\institute{Department of Physics, University of Otago\\
Box 56, Dunedin, NEW ZEALAND.} \\
\institute{Department of Computer Science, University of Auckland\\
Private Bag 92019, Auckland, NEW ZEALAND.} \\
}

\abstract{We present an update on the NZ-wide advances in the field
of Radio Astronomy \& Radio Engineering with a particular focus on
contributions, not thus reported elsewhere, which hope to either directly
or indirectly contribute to New Zealand's engagement with the international 
Square Kilometre Array (SKA) project. We discuss the status of the SKA
project in New Zealand with particular reference to activities of the 
New Zealand Square Kilometre Array Research and Development Consortium.}


\maketitle

\section{Introduction}

The Square Kilometre Array\footnote[1]{www.skatelescope.org} (SKA) is a 
billion dollar, multinational project which aims to construct the world's 
largest radio telescope\cite{Schilizzi08,Lazio09}. A consortium of 17 
countries has pledged involvement in this mega-science project which aims 
which will collect more 
information on the radio sky in the first seven hours of operation than 
has been obtained in the first 70 years of radio astronomy. The SKA will 
be comprised of thousands of individual radio telescopes, with a total 
collecting area of one square kilometre. These telescopes will be arranged in 
stations of 20 - 45 antennas. The stations will stretch over thousands of 
kilometres in order to provide resolution of the order of milliarcseconds 
which will be comparable to the best optical and infrared instruments. 
The array will be sited in either Australasia or Southern Africa and bids
to host the telescope are being coordinated by South Africa and Australia 
with both countries committed to building `pathfinder' telescopes on the
sites of the proposed SKA core in each country. 

Realization of the truly paradigm-shifting nature of the SKA will require 
significant challenges in radio engineering, data transport \& storage,
signal processing and power generation to be overcome. Much progress has
been made on many of these fronts but there is much still to do and even
seemingly small communities can play a vital role in the project. In this 
paper we describe the current status of the project in New Zealand
with particular emphasis on the research enagement from Aotearoa.

\section{SKA in Australia}
Australian researchers have been involved in the SKA project since its 
inception over a decade ago. Radio astronomy is the largest single 
discipline in 
Astrophysics in Australia accounting for over 30 per cent of the total 
astronomical research output in Australia 
\cite{decrev06}. Radio astronomy research groups currently exist in seven 
Australian universities in addition to the government funded CSIRO division
of the Australia Telescope National Facility and radio astronomy accounted for
nearly 40 per cent of all permanent jobs in Australian Astronomy
in 2005 \cite{decrev06}. With the advances toward the SKA, this fraction is 
likely to expand significantly.

To date the Australian Federal Government has committed 125 million AUD for 
construction of their pathfinder telescope, the Australian Square Kilometre 
Array Pathfinder (ASKAP) and 80 million AUD for a High Performance 
Computing (HPC) facility in Perth\footnote[2]{www.ska.gov.au/news/Pages/MinisterCarrnamesPawseyHPCCentre.aspx}. In addition, the Western Australian 
Government has committed a further 20 million AUD to establish the 
International Centre for Radio Astronomy Research (ICRAR)\footnote[3]
{www.icrar.org} as a joint venture between the state and the University of 
Western Australia and Curtin University of Technology.

\section{SKA in New Zealand}
In August 2009 New Zealand and Australia signed an arrangement 
to collaborate on the bid to host the SKA \footnote[4]{'Arrangement between
the Australian Government and the New Zealand Government on a Joint Bid
to Secure the Siting of the Square Kilometre Array in Australia and New
Zealand'}. This marked the first formal step of New Zealand's engagement
with the project and will see a joint trans-Tasman effort to work together 
to co-host the telescope.

\subsection{NZ SKA Project Office}
The Ministry of Economic Development (MED) is the lead government 
agency for New Zealand's involvement in the SKA.  A New Zealand SKA 
project office has been established as New Zealand's point of contact 
for SKA matters and is coordinating New Zealand's Government level 
programme for engagment in the SKA project. As part of the coordination effort
the project office is working closely with the Australia-New Zealand SKA 
Coordinating Committee (ANZSCC) on which the NZ Government has a 
representative.

\subsection{SKA Research \& Development Consortium}
In June 2009, just prior to the announcement of the formal arrangement
with Australia, the New Zealand Square Kilometre Array Research \& 
Development Consortium (SKARD) was formed\footnote[5]{www.ska.ac.nz}. The 
role of SKARD is to bring 
together professional researchers involved in SKA related research in NZ, 
to foster collaboration both within NZ and internationally and to liaise 
with industry, government and other groups to advance New Zealand's 
contribution to the SKA. Members are drawn from all of New Zealand's 
major research universities and have interests in antenna design, signal 
processing, imaging and inference, high performance computing and 
radio astronomy. 

\subsection{NZ SKA Industry Consortium}
In addition to SKARD, the New Zealand SKA Industry Consortium 
(NZ SKAIC) has formed to achieve positive economic 
outcomes for New Zealand from involvement in the SKA project.
The group consists of industry partners representing
IT software, hardware, networks and services, the Ministry of Economic 
Development (MED) and New Zealand Trade and Enterprise (NZTE).

\section{Research Engagement in New Zealand}
Although SKARD acts as a coordination point for researchers engaged in
SKA related research, research engagement at the individual level has
been quite active over 2009 even before the formation of the consortium. 
Several groups from across the country 
have come together to form collaborative projects around radio astronomy and 
related engineering. Highlights of this collaborative work include efforts
in radio astronomy, radio engineering and signal processing.

\subsection{Radio Astronomy}
Currently there are only two universities in which radio astronomy 
research is conducted in New Zealand.
Consequently, direct involvement in pure radio astronomy research is limited
\footnote[6]{Refereed research publications in radio astronomy for NZ 
based researchers in the last decade are confined to only two authors: 
Johnston-Hollitt \& Budding.}, however where such engagement has occurred 
it has been well placed. 
In particular, NZ has good representation on forthcoming major science
and associated technical research associated with the Australian SKA 
Pathfinder, pan-NZ capability building in radio astronomy, signal processing
and engineering and an emerging interest in associated High Performance 
Computing (HPC).

\subsubsection{ASKAP Surveys}
Australia has committed to build a one per cent demonstrator for the SKA called
the Australian Square Kilometre Array Pathfinder (ASKAP). ASKAP, which is in 
construction now and due for completion in 2012, will be a powerful new 
telescope in its own right. Located at the core of the Australian \& 
New Zealand site for the SKA, the Murchison Radio Observatory in Western 
Australia, ASKAP will 
comprise 36 15 meter diameter radio dishes spread over a 6 km diameter area. 
The configuration is designed to optimize sensitivity on angular scales of 
30 arcseconds, whilst still providing good low surface brightness sensitivity 
at lower resolutions \cite{Gupta08}. ASKAP will operate over 700 - 
1800 MHz with a one degree field of view and has been specifically designed as 
a survey instrument for redshifted observations of neutral hydrogen (HI) 
\cite{Staveley-Smith08}. ASKAP will act as a survey instrument for 
approximately 75 per cent of the first five years of operation 
(2013 - 2018) and an open call for survey science proposals in late 2008 
resulted in 38
expressions of interest from world-wide teams of researchers. An extensive,
multi-stage process of international peer review to prioritize the 
science surveys has just concluded with eight science surveys and two 
strategic programs selected\cite{Ball09}.

Several NZ-based researchers are involved in the successful surveys which were
selected for ASKAP. Staff from the Victoria University of Wellington (VUW)
are involved in three ASKAP proposals, the A ranked Evolutionary Map of the 
Universe (EMU)\cite{Norris09} and the A- ranked surveys 
``POSSUM''and ``VAST''. Staff from the Auckland University of Technology are 
part of the strategic project ``The high resolution component of ASKAP: 
meeting the long baseline specifications for the SKA''.

Within the commensally observed EMU and POSSUM projects, researchers at 
VUW will contribute to science related to clusters of galaxies. In particular, 
understanding the role of environment on the production of radio emission
in clusters will be investigated. This will include: the detection and 
characterization of low surface brightness diffuse synchrotron
emission associated with dynamical encounters \cite{mjh03}, 
the use of tailed radio galaxies as barometers of ``cluster weather'' 
\cite{mjh04} and as probes of high density regions \cite{Mao09a, Mao09b}, 
and an investigation into the properties of AGN and starforming galaxies in 
different cluster environments \cite{Venturi01, Venturi02, mjh08}.
The latter two projects are relatively straightforward in terms of finding
and identifying sources. Detection of diffuse, low surface brightness 
sources however will be a challenge and is likely to require new signal
processing techniques. Consequently researchers from VUW are undertaking
a broader program of research in signal processing for astronomical
research to address the challenge (see \cite{Hollitt09}).

The particular interest of NZ members of the VAST team is investigating 
coherent emission from active stars and binary systems. Such observations will 
probe fundamental parameters of stellar magnetospheres such as magnetic 
field intensity and topology and electron energy distribution and provide
vital information on the relation of this emission to the physical
characteristics of the host stars.

\subsubsection{Radio Telescopes in New Zealand}
The have been a range of substantial low-frequency arrays and small 
higher frequency ($\nu \leq$ 3 GHz) dishes in NZ over the last 
50 years. In particular, there has been a long history of polytechnics
associated with small dishes including a 10m dish at Hamilton built
in 1985 by the Waikato Institute of Technology in association with
Waikato University\cite{Head06} and a 5m dish constructed in 1998 by 
the Central Institute of Technology in Upper Hutt \cite{Budding00}. 
(See \cite{Head06} and references therein for a full discussion of
the history of small radio dishes in NZ.) The most recent in this 
long line of small dishes is the AUT 12m dish at 
Warkwork, located just north of Auckland \cite{Gulyaev09}. Although 
this dish, which has been constructed but is currently still 
undergoing testing, 
will primarily undertake geodetic observations \cite{Gulyaev09}, it is 
hoped it will contribute to science observations of bright radio 
sources through the Australian Long Baseline Array (LBA). 

Additionally, there are several low-frequency arrays operating in NZ for the
purposes of undertaking upper atmospheric or meteor detection research, 
which are not connected to radio astronomy research.

\begin{figure}
\includegraphics[width=\linewidth]{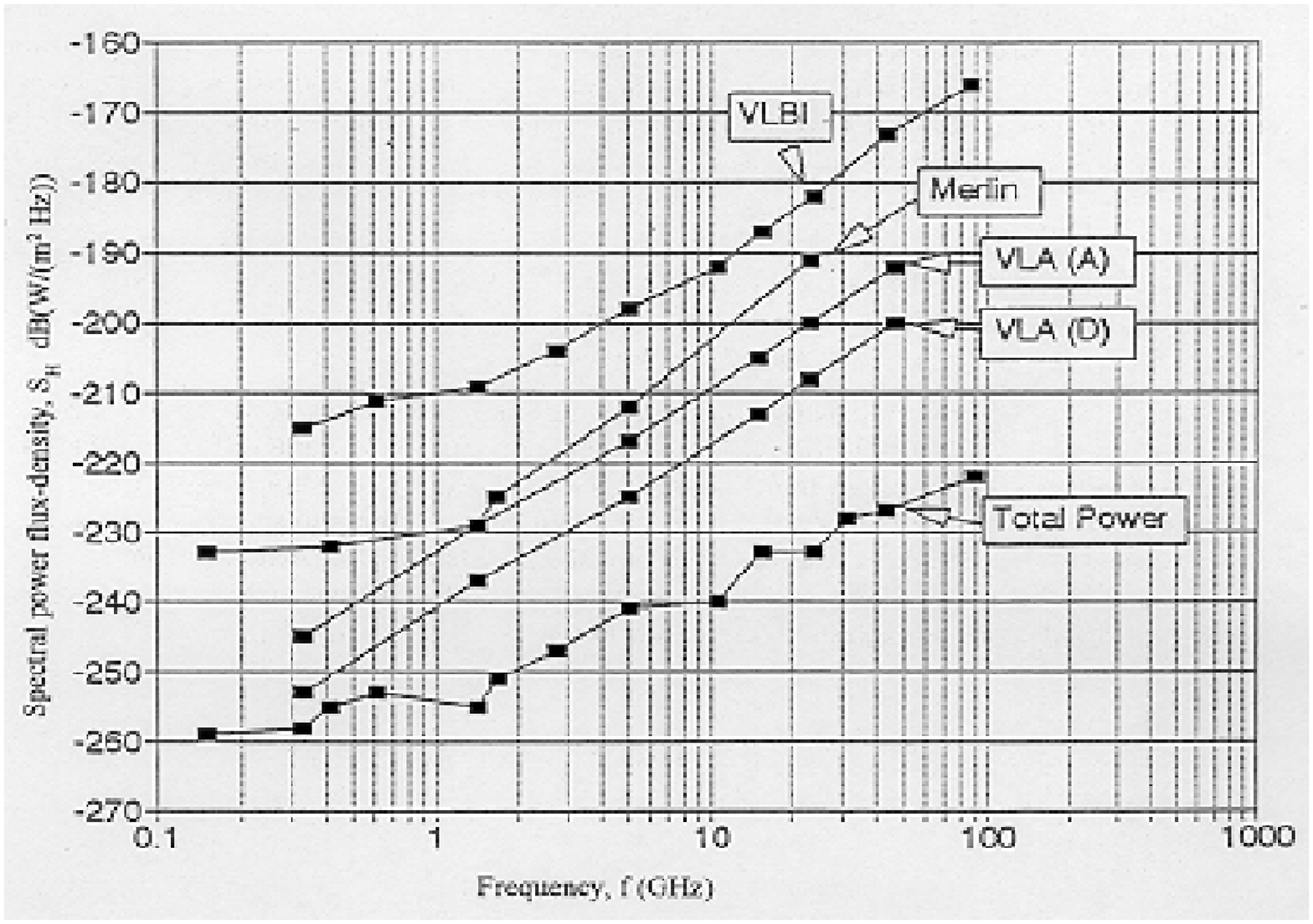}
\caption{Interference thresholds as a function of frequency for continuum
observations with several different configurations of radio telescopes. 
The line for total power describes the spectral power flux density (dB(
W m$^{-2} Hz^{-1}$)) for a single dish radio telescope. Subsequent lines
give the threshold for harmful interference for different configurations
of interferometers going from more compact (VLA(D)) to least compact 
(Merlin) configurations. The least stringent requirements are the present
recommendations for Very Long Baseline Interferometry (VLBI). Taken from
the International Telecommunication Union Handbook on Radio Astronomy
\cite{ITUHandbook}. The outer stations of the SKA, such as those 
potentially in New Zealand, will be required to conform with VLBI levels
\cite{ITURegs,SKAmemo73}.}
\label{rfi-tab}
\end{figure}

\subsubsection{Low Frequency Transient Network Development}
\label{tread}
Funding has recently been obtained to build a low frequency receiver 
network deployed across NZ to investigate properties of the transient radio 
sky. The project Transient Radio Emission Array Detector 
Prototype (TREAD-P) \cite{Kitaev09} will employ FPGA devices connected across 
the Kiwi Advanced Research and Education Network (KAREN) to test collection 
and storage of very high data rates and novel techniques for data processing 
using Bayesian inference with the hope of characterizing the low frequency 
radio sky. The core of the detector network will be the Digital Receiver 
Sensor (DRS) which is based around Field Programmable Gate Array (FPGA) 
technology, which is becoming increasingly important to radio astronomy. 
Depending on the configuration, each DRS will produce up to 10MB of data 
per second and the network of 10 devices will produce more than 8.6 Terabytes 
daily when running at maximum capacity. 

Major international radio telescopes like the Low Frequency Array (LOFAR) 
\cite{Huub03} and ASKAP have identified characterization of the transient 
radio sky as a high science priority and proposed instruments like the 
Long Wavelength Array plan to conduct science specifically on the transient 
radio sky (20 - 80 MHz) in the next decade. Instruments like the 
DRS will provide vital information on the frequency of both artificial and 
naturally occurring radio signals which will be vital to the survey design 
of these large low-frequency telescopes.

In order to achieve this goal researchers from the Auckland 
University of Technology, University of Otago, Victoria University of 
Wellington, University of Auckland and the University of Canterbury and 
industry partners have agreed to contribute resources. It is expected 
this will provide opportunities for at least three research students at 
partner institutions across New Zealand in 2010. 

\subsection{HPC Facilities}
New Zealand has capability in High Performance Computing, with dedicated 
facilities at the University of Canterbury (BlueFern) and computation 
clusters in several universities and government research labs. Many of 
these facilities are coordinated via its national grid initiative 
BeSTGRID led by the University of Auckland and seven other member 
institutions, which operates over a 10 Gb/s national research network the 
Kiwi Advanced Research and Education Network (KAREN), providing essential 
services for collaboration, computation, and data management.

NZ-based researchers are currently utilizing these tools to 
develop capability in relation to the SKA. As an example, the low 
frequency transient project (Section 
\ref{tread}) involving researchers from five institutions 
has been recently funded to exploit this infrastructure to transport, store 
and process data at high rates from a network of GPS-synchronized low 
frequency antennas across NZ for radio astronomy research.

Additionally, researchers at three of NZs universities (Otago, Auckland and 
AUT) have established a collaboration developing new high performance 
computing techniques for processing radio astronomy data with a specific 
focus on reconfigurable computing. AUT, Otago and VUW maintain HPC clusters 
and in addition AUT and Otago have reconfigurable FPGA based systems 
(based in physics and engineering respectively) while BeSTGrid (Auckland) 
provides an ideal middleware platform to develop techniques for distributed 
computations which will be required for SKA research.

While this is a foundation, it should be noted however that host countries 
for the SKA will require significant upgrades to e-Infrastructure including 
a distributed high-speed fibre optic network between the stations and the 
science processing facility, mega-watt power for SKA stations and 
specialised hardware for the central digital signal processing facility. 
Depending on how the SKA is configured many of these 
resources may be located in NZ as part of the international project. 

\section{Discussion: Immediate goals}

\subsection{Station Locations}
In partnering with Australia to host the SKA, New Zealand is hoping to
contribute the outer stations of the array. At this stage this is likely
to mean two stations of antennas in NZ, though arguments for more stations
could be made in the future. In order to achieve the science
goals of the SKA even the outer stations of the array will require low
levels of radio frequency interference (RFI)\cite{SKAmemo73}. 
Figure \ref{rfi-tab} 
shows the currently required levels specified for outer stations of the 
SKA in the low to mid frequency bands\cite{ITURegs}. While there have been
a couple of suggested locations put forward for SKA stations none of these
meet the required specifications.

\begin{figure}
\includegraphics[width=\linewidth]{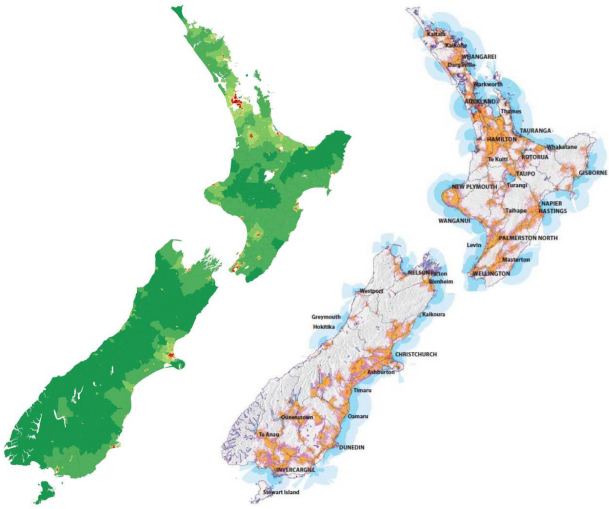}
\caption{LHS: Population density of NZ from the 2006 Census$^7$; RHS current 
mobile phone coverage for NZ$^8$. To reduce the levels of 
RFI, potential SKA stations should ideally be located in regions of low 
population density with little mobile coverage.}
\label{census}
\end{figure}

It is probably worth noting here that the vast majority of RFI is
man-made and easiest way to achieve the radio quite levels required for the
SKA is to place the antennas in regions of low population density. Figure
\ref{census} shows the current population density of NZ as determined from
the 2006 national census\footnote[7]
{http://en.wikipedia.org/wiki/File:NewZealandPopulationDensity.png} in 
addition to Telecom's current mobile phone 
coverage\footnote[8]{www.telecom.co.nz}. Regions of low population density and no mobile coverage exist in
much of the South Island and there are some pockets in the North Island with
similar characteristics. 

At present, a comprehensive and independent site selection and survey 
process in close collaboration with Australia is currently being 
negotiated through the NZ SKA Project Office. As with all telescope site
selection processes this will examine a range of metrics before finalising 
potential sites.

\subsection{Research Engagement}
With the signing of the arrangement with Australia and formation of SKARD and 
NZSKAIC, New Zealand is now on a firm footing for researchers from the private
and public sector to engage in this project. As the final host nations
for the array will not be decided until the early part of 2012 it is important
for NZ to maximise benefit in the project via early engagement while 
simultaneously mitigating the risk that the project will go to Southern 
Africa. This means that it is important to identify niche opportunities 
which are
independent of the ultimate location of the array. 
This can be achieved via expanding existing links between NZ and 
Australia in radio astronomy related areas via:\\
i)	continued participation of NZ scientists in the science and design 
of the SKA,\\
ii)	building expertise in NZ through the training of graduate students 
in radio astronomical research - a natural complement to NZ's existing 
strength in optical astronomy, and\\
iii)	undertaking significant and ground breaking research which will be 
used for further enhancement of science with next generation radio telescopes.

The latter speaks directly to the identification and exploitation of niche 
research opportunities. Such niches have clearly
emerged in NZ around science-driven post-processing via novel signal processing
\cite{Hollitt09}. Work of this nature combines NZ's strengths in radio 
astronomy, signal processing, HPC \& middleware. Additionally, capacity 
building exercises and pan-NZ collaborative work \cite{Kitaev09} which has 
commenced will be 
important to realize opportunities that will arise with hosting the telescope.

On the purely astronomical research front, in order to maximize the 
science outcomes of both ASKAP and eventually the SKA it is crucial that 
knowledge from existing instruments is examined and used to developed 
expertise vital for survey design on next generation instruments and the 
research teams thus far involved are well placed to make significant 
contributions.

\section*{Acknowledgements}
MJ-H acknowledges continuing support towards radio astronomy research by the
Victoria University of Wellington.


\bibliography{enzcon}

\end{document}